# MPI Streams for HPC Applications

Ivy Bo PENG [a], Stefano MARKIDIS [a,1], Roberto GIOIOSA [b], Gokcen KESTOR [b] and Erwin LAURE [a]

[a] *Computational Science and Technology Department, KTH Royal Institute of Technology, Stockholm, Sweden*
[b] *Computational Science and Mathematics Division, Pacific Northwest National Laboratory, WA, USA*

**Abstract.** Data streams are a sequence of data flowing between source and destination processes. Streaming is widely used for signal, image and video processing for its efficiency in pipelining and effectiveness in reducing demand for memory. The goal of this work is to extend the use of data streams to support both conventional scientific applications and emerging data analytic applications running on HPC platforms. We introduce an extension called *MPIStream* to the de-facto programming standard on HPC, MPI. MPIStream supports data streams either within a single application or among multiple applications. We present three use cases using MPI streams in HPC applications together with their parallel performance. We show the convenience of using MPI streams to support the needs from both traditional HPC and emerging data analytics applications running on supercomputers.

## 1. Introduction

Streaming computing is a programming paradigm that supports reactive real-time computation on irregular, potentially infinite, data flow [1]. Streams are a continuous sequence of fine-grained data from a set of processes, called data *producers*, to another set of processes, called data *consumers*. The computation on each data stream is on-line as data streams are discarded (or *consumed*) after the data has been processed. Streaming computing is widely used in image and video processing, where data is continuously streamed and processed on-the-fly. Streams are also used to support real-time computing that requires a prompt reaction to an event. Currently, streams are not widely used in High-Performance Computing (HPC) for different reasons.

The first reason is that many conventional HPC applications have regular coarse-grained communication pattern that is not a good fit for streaming computing. In fact, HPC applications usually aggregate several small messages into a larger message to avoid the fixed communication overhead (latency) of sending multiple messages. On the contrary, data streams are typically fine-grained and irregularly communicated. However, new HPC applications have emerged in the last decade with demands for both streaming and computing capabilities on supercomputers. Examples of such applications are the applications that require the on-line processing of large data sets from experiments [2], such as Large Hadron Collider (LHC) [3], Square Kilometer Array (SKA) [4], Large Synoptic Survey Telescope (LLST) [5] and the Laser Interferometer Gravitational-Wave Observatory (LIGO) [6]. All these infrastructures have provided high data rate and will provide even higher data rates. SKA[4] is estimated to reach 10 PB/s, which can drain the entire memory of supercomputers if data is not being filtered in time. To handle such large volume of data, it requires on-the-fly extraction of interesting data (signal) and concurrently disregarding the uninteresting data (background), which is the essence of streaming computing.

The second obstacle for adopting streaming computing on supercomputers is the lack of support in popular programming systems. For instance, MPI is the de-facto approach for parallel programming on distributed-memory systems like supercomputers. However, it does not and will not provide (at least in the next MPI-4) functions to support streaming computing. A series of approaches, such as Pebbles [7] and one-sided MPI communication with notification [8], can in principle be conveniently adapted to support streaming computing. However, no explicit support has been provided. Recently, we have extended MPI to support streaming operations by implementing a library, called *MPIStream* [9]. This library allows us to divide MPI processes into groups of data producers and consumers and connect them with asynchronous and irregular data streams. In this paper, we demonstrate concrete use cases of streaming computing in scientific and data analytics applications on supercomputers.

The goal of this paper is to advocate the use of streaming computing in HPC applications. We first show the use of streaming in a single application that follows the MapReduce model. A group of processes are dedicated to map tasks while a second group of processes are dedicated to reduce tasks. These different groups are connected via MPI streams. A second example is to use MPI streams to couple two separate MPI-based applications. A particle-in-cell code carries out plasma simulations and streams out particle data to another application that performs irregular high-frequency I/O operations and visualization. These two applications are coupled with MPI streams at runtime. This technique allows for decoupling the I/O operations (potentially in-situ visualization) from the scientific application. Finally, we mimic the LHC experiment continuously streaming particle collision data to a set of MPI processes that classify the collision events and capture only the events of interest.

The paper is organized as follows. We first give an overview of the previous work on streaming computing in Section 2. A description of MPI streams and related concepts follows in Section 3. We show three examples of using MPI streams in HPC applications with their performance results in Section 5. Finally, we summarize and discuss the results in Section 6.

## 2. Related Work

The theoretical framework of streaming computing stems from the work on data flow [10] and Khan process networks (KPN) [1]. KPN provides a simple parallel computing model that connects multiple processes with first-in-first-out channels with flexibility in constructing the process graph. KPN is a semantic model providing formal proof of various properties such as determinism and termination. In KPN, streaming computation is viewed as a continuous function and the streaming computing is determined by calculating the least fixed point of this function [11]. Recent works have proposed KPN as a more flexible data processing model compared over other data processing frameworks [12]. However, the KPN implementations are often limited by the memory consumption and difficulties arise in establishing communication channels and scheduling tasks. In fact, these issues impose constraints to some properties in KPN. Extensive works have been carried out for either mitigating such limitations or extending flexibilities in implementations [13,14].

Several streaming languages have been implemented for the development of stream-based applications [1]. StreaMIT [15] is one modern streaming language providing an high-level abstraction of streaming model. StreaMIT takes advantage of `Java`-based compiler for delivering high performance computation. StreaMIT constructs programs using *Filter* as the basic computation unit and connects these filters using data streams as the basic communication unit. Based on the assumption of static data flow rate, a timing mechanism that is relative to the data flow is provided in this language to facilitate irregular control messages. Despite the success of StreaMIT on other platforms, there are several limitations that prohibit it from being adopted on supercomputers. First, it is very difficult to reformat most existing HPC applications completely complying to the stream abstraction. Instead, for easier adoption by the exiting HPC applications, it is necessary to provide a natural interface to these current applications. Second, the de-facto programming systems on HPC is MPI and its most active implementations are in `C` and `Fortran`. Third, most streaming languages are designed for productivity instead of performance. Because performance is not the main goal, these approaches underutilize the computing resources from HPC platforms.

Streaming computing has been adopted for signal processing, image rendering and media processing. In these domains, special architectures and processors have been designed to accelerated the execution of applications formatted in the streaming models. Typically, these streaming processors are based on VLSI technology that utilizes a large number of ALU, stream register files and high data bandwidth to exploit the natural data parallelism and pipelining in these domains of applications [16,17,18,19]. These specialized hardware can greatly boost the performance of streaming applications. However, their specialization also means that only a small portion of applications can benefit from such special hardware, limiting their scales as other general-purpose systems.

Today large datasets are often processed by cloud computing for their widely accessible resources and for economic reasons. MapReduce [20] is the dominant programming model on cloud computing [21]. The MapReduce model requires data to be formatted as key-value pairs, limiting the variety of possible data computations. Besides the MapReduce model, streaming frameworks have also been developed for cloud computing, such as Apache Spark [22]. These frameworks are primarily designed for batch processing data from the file systems. Thus they are designed for high throughput to the file systems instead of low latency of a single data access. In fact, many of these frameworks use the file systems to store job states and intermediate results. Data is first captured for an interval and then processed in batches, resulting in high latency and overhead. Also these frameworks mainly focus on providing fault-tolerance because of the cloud setting where failures and faults occur frequently. These design objectives do not match the characteristics of use cases on supercomputers, where the data is generated by simulation applications at runtime and needs to be processed with low latency without involving considerably the file systems. In addition, as these frameworks aim for coding productivity, their APIs are often available in `Java`, `Scala` and `Python` instead of the dominant programming languages for supercomputers `C` and `Fortran`.

From recent observations, more and more data-intensive applications are moving to the HPC platforms for taking advantage of either larger memory capacities or high processing rates. This has motivated works on extending the de-facto programming system on supercomputers, MPI, to support the streaming model. MPIStream is one prototype library that was implemented atop MPI and provides an interface to existing MPI applications to partially or fully adopt the streaming model [9]. A further study on the impact factors of streaming computing on supercomputers has been reported in [23], characterizing the performance of streaming computing by the process topology, injection rate and processing rate. Streaming computing on supercomputers imposes many different challenges compared to other platforms. For instance, the data source on supercomputers could be large-scale experiments, such as LHC, or the results of large-scale simulations. In several cases, the data needs to be processed in real time, streaming directly to data analytics processes without involving filesystems because of both performance and energy considerations. Also, the data producer and consumer applications are often independent programs running on different processors or even different compute nodes. Thus, instead of integrating both applications into one application, it is more flexible and efficient to provide a streaming coupling capability to connect these applications at runtime with MPI streams.

## 3. MPI Streaming Computing

In this section, we introduce the streaming model and its implementation in an MPI-Stream library.

### 3.1. MPI Streaming Model

Our MPI streaming model is based on four basic concepts:

1. **MPI channel.** An MPI channel is a persistent connection between two groups of processes. A channel can support different data streams. Data producers and consumers collectively construct a communication channel by calling the function `MPIStream_CreateChannel()`. Each process indicates their role as either a data producer or a data consumer in the newly created channel.
2. **MPI data producer and consumer communicators.** These two communicators are set up after a channel is established so that MPI communication among data producers can be isolated from MPI communications among data consumers. When coupling multiple applications, these two communicators will be used as the initial communicator to replace the default `MPI_COMM_WORLD`. We use data producer and consumer processes to refer to those processes belonging to each respective communicator.
3. **MPI stream.** An MPI stream is a continuous, irregular flow of data in the form of basic or derived MPI datatypes [9] between data producer and consumers. An MPI stream is associated to an MPI channel. The structure of the data streams flowing in this channel will be defined by an MPI (derived) datatype. This definition of stream element needs to be consistent on both data producers and consumers as it is the basic coupling unit between the two.

**Table 1.** The six basic functions of the MPIStream library and their description. The third column indicates whether the function can be only used on producer (P) or consumer (C) processes or both (P-C).

| MPIStream Function | Description | P/C |
|---|---|---|
| int `MPIStream_CreateChannel`(int isProducer, int isConsumer, MPI_Comm comm, MPIStream_Channel* channel) | This is a collective operation called by all MPI processes in MPI_Comm comm. The MPIStream_Channel object is set up on processes that are data producer or consumer. | P-C |
| int `MPIStream_FreeChannel`( MPIStream_Channel* channel) | This operation releases all the resources allocated for the communication channel. | P-C |
| int `MPIStream_Attach`(MPI_Datatype streamDataType, MPIStream_Operation operation, MPI_Stream *stream, MPIStream_Channel *channel) | This function defines the stream data structure to be attached to a given communication channel as well as the logics for processing each data stream. It sets up an MPIStream object on data producers and consumer processes. | P-C |
| int `MPIStream_Send`(void* sendbuf, MPI_Stream *stream) | This function is called by a data producer to stream out one stream element belonging to the specified stream. This function includes blocking and non-blocking versions for different use cases. | P |
| int `MPIStream_Terminate`(MPI_Stream* stream) | This function is called by a data producer to signal the termination of its contribution to the specified stream. Note that data consumers will only consider a stream terminated when all of its data producers have signaled the termination. | P |
| int `MPIStream_Operate`(MPI_Stream* stream) | The function is called by a data consumer to receive and process the incoming stream elements. This function includes blocking and non-blocking versions for different use cases. | C |

4. **MPI stream operation.** Each MPI stream has an attached operation specifying how each stream element should be processed once received by a data consumer. The defined structure of data streams and the functions used for processing the data are attached to a communication channel by calling the function `MPIStream_Attach()`. Data consumer processes also need to specify the data processing function, as well as (optional) initialization and terminal functions by providing callback functions to the MPIStream library.

The basic functions in the MPIStream library are listed in Table 1. The integer return value of the six functions provides error codes indicating if the function executes correctly. The last column indicates whether a function is called by a data consumer or by a data producer process. These functions include blocking (function does not return until the communication is finished) and non-blocking (function returns immediately) versions for different use cases, i.e. `MPIStream_Isend` is available together with `MPIStream_Send`.

The MPIStream library implementation is based on MPI persistent point-to-point communication. Details of the implementations and performance results on a modified STREAM benchmark are provided in [9].

*3.2. Using the MPIStream Library*

MPIStream can be used within a single application to define MPI process as data producers and /or consumers and connect them by MPI streams. This mode is often useful when there are some irregular operations in a parallel application that can be decoupled to a smaller group of processes. This could be very efficient to decouple collective operations that are called to determine dynamic and irregular workload on all processes. In this paper, we show an example of decoupling computation and irregular high-frequency I/O operation to allow runtime visualization.

A second way to use MPIStream is to couple two or more separate applications. In this way, each application can act as either a data producer or consumer or both. In addition to their usual simulation/operations, the applications acting as data producer can stream out data to other applications as early as partial result is ready. For example, matrix multiplication program can stream out data as early as slices of a three-dimensional matrix have been calculated and can be used by another application. To enable this mode, all applications need to be launched in multi-program multi-data (MPMD) with job scheduler. This second approach is very relevant for coupling different codes solving different physical models to realize distributed multi-physics computational framework. For instance, when simulating space plasmas in planetary magnetospheres, plasma kinetic models that are computational expensive can be solved in small spatial regions (where kinetic effects are important) and coupled with a fluid model that provides the global evolution of the system [24]. Such capability requires different parallel programs to exchange data and this can be achieved by using MPI streams. Another example is the coupling of an HPC application with an in-situ data analytics and visualization application. In this case a main application carries on computations and streams out data asynchronously to an application that performs on-line analysis.

We demonstrate the second usage in Listing 1, showing the skeleton code for coupling a data producer application on the left panel (*a*) and a data consumer application on the right panel (*b*). During initialization of the applications, all the processes create a communication channel with `MPIStream_CreateChannel` and define a parallel stream in that channel with `MPIStream_Attach` The basic stream element is an array of 10 `MPI_INT`, defined as an MPI derived data type (`MPI_Type_contiguous`), in the example of Listing 1. During the execution, the data producer application continues streaming out data stream elements with `MPIStream_Send` while the data consumer application performs computation on the incoming stream elements with `MPIStream_Operate`. The operation to be performed on the MPI streams are defined on the data consumer application as call-back functions (`setup` is to initialize the computation, `onlineFilter` is the actual operation on the stream and `finalize` is used to finalize the computation in Listing 1). When a data producer finishes sending all its data, it signals the termination of the stream with `MPIStream_Terminate`. The data consumers continue receiving and processing the incoming stream data until the corresponding data producers communicate the termination of the stream.

```c
#include "MPIStream.h"
int main(int argc, char** argv)
{
    int myrank, nprocs;
    MPI_Comm my_comm;
    MPI_Init(&argc, &argv);

    //indicate the process is a data producer
    int is_data_producer = 1;
    int is_data_consumer = 0;

    // 1. establish a communication channel
    MPIStream_Channel channel;
    MPIStream_CreateChannel(is_data_producer,
                            is_data_consumer,
                            MPI_COMM_WORLD,
                            &channel);
    myrank  = channel.ProducerRank;
    my_comm = channel.StreamProducerComm;
    nprocs  = channel.ProducerSize;

    // 2. specify the structure of data streams
    MPI_Datatype streamDatatype;
    MPI_Type_contiguous(10,MPI_INT,&streamDatatype);
    MPI_Type_commit(&streamDatatype);

    // 3. attach the specified operation to a
    //    parallel stream
    //    null operation on data producer
    MPIStream stream;
    MPIStream_Attach(streamDatatype,
                     NULL,
                     &stream,&channel);

    //4. Start the usual simulation
    while(!done){
        ...//computation
        ...
        ...
        if(needed)
            MPIStream_Send(&data,&stream);
    }

    //5. Finalize
    MPIStream_Terminate(&stream);
    MPIStream_FreeChannel(&channel);
    MPI_Finalize();
    return 0;
}
```

(a) MPI data producer application.

```c
#include "MPIStream.h"
// user-defined operations on streams
void setup()
{ ... }
void onlineFilter(void *in)
{ ... }
void finalize()
{ ... }
int main(int argc, char** argv)
{
    int myrank, nprocs;
    MPI_Comm my_comm;
    MPI_Init(&argc, &argv);

    //indicate the process is a data producer
    int is_data_producer = 0;
    int is_data_consumer =1;

    // 1. establish a communication channel
    MPIStream_Channel channel;
    MPIStream_CreateChannel(is_data_producer,
                            is_data_consumer,
                            MPI_COMM_WORLD,
                            &channel);
    myrank  = channel.ConsumerRank;
    my_comm = channel.StreamConsumerComm;
    nprocs  = channel.ConsumerSize;

    // 2. specify the structure of data streams
    MPI_Datatype streamDatatype;
    MPI_Type_contiguous(10,MPI_INT,&streamDatatype);
    MPI_Type_commit(&streamDatatype);

    // 3. specify the data processing routines
    MPIStream_Operation operation;
    operation.init_func = &setup;
    operation.proc_func = &onlineFilter;
    operation.term_func = &finalize;
    operation.bg_func   = NULL;

    // 4. attach the specified operation to a
    //    parallel stream
    MPIStream stream;
    MPIStream_Attach(streamDatatype,
                     &operation,
                     &stream,&channel);

    // 5. start processing the data
    MPIStream_Operate(&stream);

    // 6. Finalize
    MPIStream_FreeChannel(&channel);
    MPI_Finalize();
    return 0;
}
```

(b) MPI data consumer application.

**Listing 1** Coupling data producer and consumer applications with the MPIStream library in MPMD fashion.

### 4. MPI Streams in HPC Applications

In this section, we demonstrate using MPI streams to enable effective pipelining and communication either between processes of a single application or among different HPC applications. In particular, we show three applications including a data analytics workload (MapReduce), a scientific HPC simulation and the analysis of large scale experiment data.

First, we use MPI streams within the same application to implement a MapReduce program. Different from previous works on implementing a MapReduce library in MPI [25], our use case demonstrates the flexibility of using MPI streams to support a popular data processing model. Second, we decouple I/O operations and visualization stages from a plasma simulation code to a separate application and connect them with MPI streams. Third, we mimic the LHC sensors by having a set of processes reading *HIGGS* data sets of particle collision events [26] and stream the event data to a data analytics application that has been trained to distinguish signal and background events and capture only signal events on-the-fly.

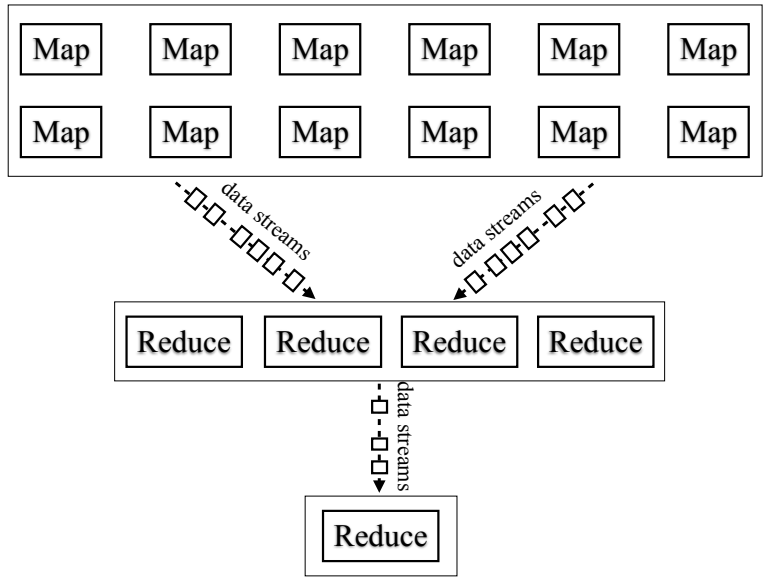

**Figure 1.** A conceptual diagram illustrating MPI streams for distributed MapReduce on supercomputers.

*4.1. Benchmark Environment*

We use the KTH Beskow supercomputer for testing the performance of the three applications using MPI streams. Beskow is a Cray XC40 supercomputer with Intel Haswell processors and Cray Aries interconnect network with Dragonfly topology. The supercomputer has a total of 1,676 compute nodes of 32 cores divided between two sockets. The operating system is Cray Linux, and the applications are compiled with the Cray C compiler version 5.2.40 with optimization flag -O3 and the Cray MPICH2 library version 7.0.4.

*4.2. MapReduce with MPI Streams*

We first show an implementation of a MapReduce program using data streams connecting the map and reduce processes. MapReduce is currently the most popular parallel programming model in processing massive datasets [20]. MapReduce is a simple model as the user only needs to specify two functions to map a given key-value map and to reduce a key and its value list. Nevertheless, it is also a very powerful model that has been proved effective for various data transformation and complex graph problems. Previous works on supporting MapReduce in MPI either provide a specific library for this model [25] or explore collective communication in MPI for reduce operation [27]. In both works, all-to-all communication is used for the global knowledge of keys and also in case of irregular key distribution. In this paper, we explore the possibility of using data streams for regularizing irregular distribution of key-value pairs for MapReduce program on supercomputers. This is achieved by linking map and reduce tasks with irregular and asynchronous data streams. The original MapReduce model has a master process that accesses data and administrates tasks to workers. By simply applying master-worker mode, it is common to incur into hotspots and congestion when scaling to a large number of workers on parallel machines. In fact, if the input data is located in the file system, it is nearly impossible to keep all workers busy if there is only one master accessing the data. In the context of this paper, we relax the constraint of one master to a group of processes that read data and build reduce inputs.

We show that an MPI streaming model can effectively enable the pipelining between map and reduce tasks with a limited requirement for memory. Figure 1 illustrates a MapReduce program on supercomputers using data streams between map and reduce tasks. In this example, a MapReduce application consists of three groups. First, one group of 12 processes retrieves data either from file system or directly from another application. This group of processes then maps each input key-value pair $<K_i, V_i>$ to output $<K_o, V_o>$ according to a user-defined map function. The input files can be very large datasets in realistic cases, e.g. webpage contents. In such case, it is possible to have less memory on system than the memory required to store the whole data sets. In addition, the reduce task can be pipelined as soon as some mapped key-value pairs are generated. Based on these two factors, a streaming MapReduce can relax the requirement on memory as the map processes can retrieve partial data, generate mapped key-values pairs, stream them out to reduce processes and then discard the data. Meanwhile, the reduce tasks can progress in parallel when the map tasks are still ongoing. In this example, there are two levels of reduce processes. In general case, there could be multiple levels of reduce processes forming a tree structure.

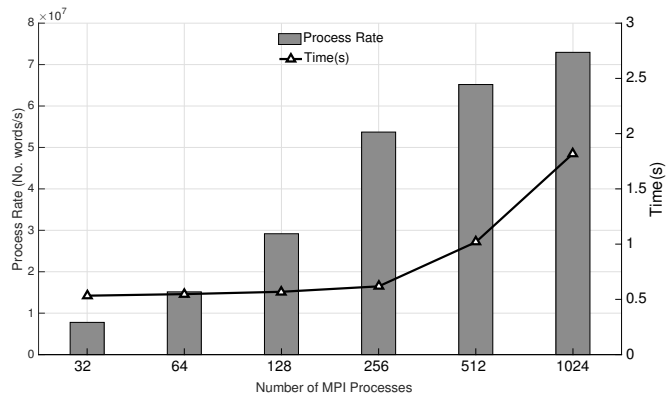

**Figure 2.** A weak scaling test of a streaming MapReduce program counting word frequency in 100 poems up to 1,024 cores on Beskow.

We evaluate the performance of the streaming MapReduce with a program that discovers the frequency of each word appearing in a collection of poems [28]. In this test case, the map function simply maps each word found in a poem to have a value of one occurrence. The reduce function then aggregates all key-value pairs with the same key by summing up their values. The final output lists out all distinct words appearing in all poems and their number of occurrences. Each map process reads from 100 poems and builds input sets from all words. The reduce processes continue receiving the key-value pairs that are streamed out by the map processes. We use the support for hash map from `C++` Standard Template Library (STL) to store and accumulate distinct keys in the streaming MapReduce code.

We conduct a weak scaling test so that the number of map processes increases and the total number of input poems also increases proportionally. The number of map processes scales from 32 to 1,024. The reduce processes are arranged into two levels with 32 processes in the intermediate level and one process in the last level respectively. The reduce process on the last level outputs the final results at the end of processing. The processing rate is calculated as the total number of words processed divided by the average execution time on each process. The experiments are repeated and have less than 5% variance in the execution time. Figure 2 presents the scaling test results of the processing rate (grey bars) and execution time (black line). The aggregated processing rate with 32 map processes, is approximately 15 million words per second. The processing rate shows nearly linear scaling when the number of map processes increases from 32 to 256 processes, reaching a processing rate of 55 million words per second. Beyond that, the processing rate continues increasing but at a slower rate. The program can process at a rate of 73 million words per second with 1,024 map processes. We note an increase in execution time on 512 processes. There could be two reasons. One reason can be that the communication between map and reduce processes is affected by congestion in network. A second reason can be that the map processes have a contention when accessing the file system. We repeated the test with 64 processes for the intermediate level of reduce processes (results not presented here) but found minimal variations in the processing rate values. Thus, we attribute the increase in execution time more to the contention when a large number of processes are accessing file systems. We note that using MPI collective I/O could improve performance when multiple processes are reading or writing to a shared file. However, in the realistic case of analyzing webpages, it is more likely that data are stored in smaller separate logs. Thus, we choose to test the more difficult case. A simple workaround will be preprocessing the files and concatenate them into a few large files and use collective I/O to mitigate contention. Overall, our test case shows that a simple streaming MapReduce program can reach high processing rate on supercomputers.

*4.3. Coupling an HPC Application with I/O and Visualization Application with MPI Streams*

The computation in HPC scientific applications can be decoupled from I/O and visualization stages by using MPI streams. In this case study, we couple a plasma simulation application, called *iPIC3D*, with a separate program that performs I/O and visualization. The iPIC3D code is a massively parallel Particle-in-Cell (PIC) code. It is used to solve space plasma physics problems [29] with applications to space weather. The simulations are an important approach for studying the solar storms to mitigate their consequence on human and technological assets in space and on the Earth. The iPIC3D code was implemented in `C++` using MPI for parallelism. iPIC3D is a highly scalable code that reaches 80.9% efficiency on 524k processes on the Mira supercomputer at Argonne National Laboratory (an IBM Blue Gene/Q machine with 49,152 nodes). iPIC3D is an open-source code and available at `https://bitbucket.org/bopkth/ipic3d-klm`.

A particle-based application, such as iPIC3D, uses computational particles to mimic plasma particles moving under the effect of electric and magnetic forces. At each computational step, the trajectory of each computational particle needs to be calculated and updated. After that, the Maxwell's equations are solved on a Cartesian grid to calculate the electric and magnetic fields. The accuracy of the simulation largely depends on the number of computational particles used for a simulation. A typical simulation on 2,000 MPI processes uses billions of computational particles [30]. These particles carry important information for understanding kinetic dynamics in plasma physics, e.g. distribution functions, phase space and particle trajectory. Thus, physicists often require the particle data to be saved for future analysis. However, the large number of particles makes the data saving a bottleneck for both the runtime performance and the post-processing because useful information is buried in the enormous amount of data. For instance, to reconstruct the trajectory of high energy particles, tens of TB of data need to be saved and processed [31].

The current production version of the iPIC3D code uses the MPI collective I/O for saving snapshots of relevant quantities, such particle positions and velocities, to disk. In this use case, we present an alternative way of using the MPIStream library to couple the simulation with a program performing I/O and visualization at runtime so that the MPI processes that carry out the simulation are isolated from the frequent and expensive I/O operations. We use the MPIStream library to decouple the processes from I/O operations by streaming out the particle data to the I/O program so that simulations can proceed without carrying out I/O operation. Concurrently, the I/O and visualization program continues processing the received particle data. A visualization of high energy particles trajectories with Paraview [32] application is presented in Figure 3. In this case, particle data is streamed out to the I/O and Paraview program at a frequency as high as each time step. Thus, no data is lost for these particles of interest and their motion can be tracked accurately.

The streaming channel between the simulation code and the I/O and visualization program is established similarly to the example in Listing 1. In this case, the iPIC3D code is the data producer and the I/O and visualization code is the data consumer. The stream element is the basic unit of the communication between these two programs. It is defined as the structure of a single particle that consist of 8 scalar values: particle position $x, y, z$, particle velocity $u, v, w$, particle charge $q$ and an identifier $id$. For tracking high energy particles, only those particles with energy exceeding certain thresholds are streamed out. It is unpredictable when and which particles will reach high energies. Thus, a particle is streamed out during particle mover, where the location and velocity of each particle is calculated. Once a particle reaches high energies, it is continuously tracked in the remaining of the simulation.

The I/O and visualization program continues receiving particle streams from the simulation at runtime and processing them to prepare data in file formats, such as VTK, that can be visualized on-the-fly by the Paraview application. The I/O and visualization program can flush data to the file system at a user-defined frequency. With sufficiently high frequency, the user can visualize the real-time motion of particles during simulation.

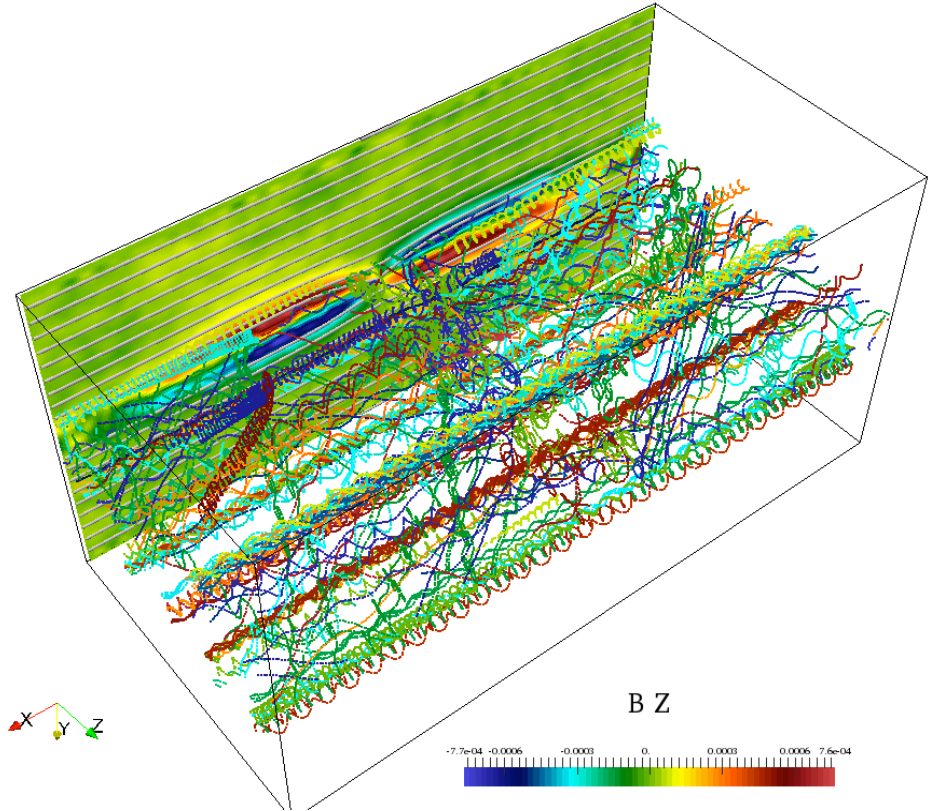

**Figure 3.** Tracking high energy particles by coupling a plasma simulation code with a visualization program with the MPIStream library. The particle trajectories are depicted in different colors for different particles, overlaying a contour plot of the z-component of the magnetic field $B_z$.

Due to the decoupling, there are four additional possibilities for the performance optimization. First, the simulation continues without waiting for the completion of any I/O operation. This provides overlapping between simulation and I/O operations. We note that MPI-3 provides non-blocking I/O operations but at the cost of additional memory buffer holding the data till the completion of such operations. This additional memory buffer might consume to much resources if the number of particles is large. Second, as the visualization program does not carry out simulation, it can stretch out most of its memory for buffering the received particles to reduce the interaction with the file system. Third, decoupling irregular I/O operations from the large number of simulation processes drastically reduces the number of processes interacting the file system, mitigating congestion and serialization for collective I/O operations. Fourth, due to the unpredictable pattern of interesting particles, it is necessary to use global reduction operations to calculate the offset of each process in a shared file. However, the communication time for such collective operations scale logarithmically with the number of calling processes. Decoupling these operations to a smaller number of I/O processes can reduce the communication time.

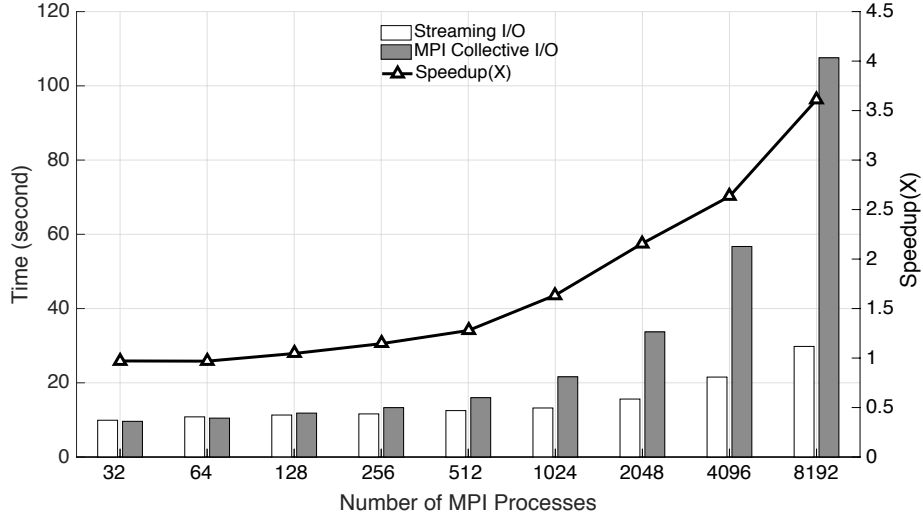

**Figure 4.** Weak scaling test of the particle visualization using MPI collective I/O and the streaming model. Each compute node is assigned with a fixed problem size. The streaming I/O uses one visualization process for every 15 simulation process.

The performance gain from the streaming model increases as the scale of the system increases. This is visible from Figure 4 that presents the scaling test results comparing the particle visualization using MPI collective I/O (in grey bars) and the streaming model for decoupling I/O from simulation (in white bars). The simulation runs for 100 time steps and the total execution time is shown in Y axis. The improvement is calculated by dividing the execution time in MPI collective I/O over the execution time in the streaming model. The improvement results are shown in the solid line against the secondary Y axis. On small number of processes, the two approaches show comparable performance. Starting from 256 processes, the streaming model demonstrates a steady improvement that continues increasing to 3.6X speedup on 8,192 processes. The observed results are in line with the fact that I/O operations are often the performance bottleneck because of hardware limitations. On supercomputers, the massive parallel programs could suffer from network contention when a large number of processes concurrently access the file system. MPI collective I/O uses a collective buffering scheme that aggregates data from all processes in one communicator and then reduces the interaction with the file systems to those aggregator processes. While this approach can effectively reduce contention and support non-contiguous data access, it is still a collective operation involving all processes. In addition, it introduces synchronization points in the the application. Processes on large parallel systems are imbalanced either because of different workload or because of system noise [33]. The cost for synchronizing the imbalanced processes could be prohibitively expensive [34] so that the applications have to reduce the frequency of I/O operations. On the other hand, the streaming model decouples the I/O operations to a separate group of processes that do not carry simulation. In this way, I/O operations and simulation are pipelined and can progress in parallel.

*4.4. Streaming Experiment Data with MPI Streams*

Particle physics uncovers the fundamental laws in the Universe. One effective approach to understand these lawas is to accelerate particles to very high energies and let them collide with each other and analyze the products of these collisions. By searching in the collision byproduct, new particles can be discovered. The LHC operated by the European Organization for Nuclear Research (CERN) is currently the largest collider in the world. During its operation in 2012, the long sought Higgs boson was discovered and confirmed. Despite the effectiveness of this approach, the LHC experiments are extremely complex and challenging. Apart from operational difficulties in physics, the enormous amount of data produced by each experiment imposes great challenges in data analytics. Millions of particle collisions could happen within one second. Each collision event can generate over ten MB data. In fact, the volume of experiment data exceeds 30 PB even after filtering out 99% events [35]. As a result, it is critical to be able to process the data fast and accurate during the experiment.

Recently, machine learning has been applied to processing the experiment data in search for new particles [36]. Given a sufficiently trained machine learning program, an input event should be classified as interesting (signal) or not (background) with high accuracy in reasonable time. Ideally, when the LHC sensor detects an event during the experiment, it sends the measurement of a list of event attributes to the classification program for real-time processing. In this way, background events could be filtered out and only signal events are captured. This can largely reduce the amount of data. However, this approach requires very high processing rate that is only available on the massively parallel supercomputers. The described program is illustrated in Figure 5.

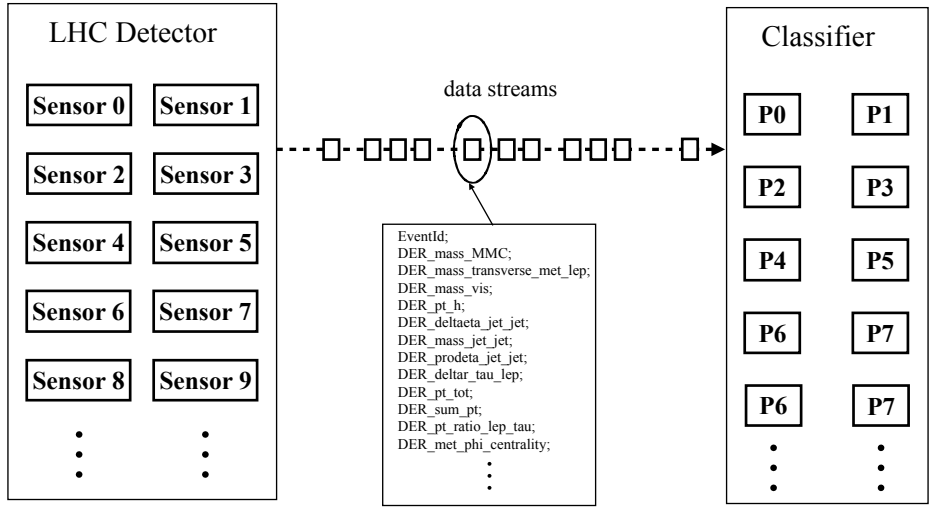

**Figure 5.** Coupling LHC detector with an event classification program on supercomputers.

In this work, we demonstrate the use of MPI streams for coupling the experiment detectors with a event classification application that runs on supercomputers. When detector captures a collision event, it measures a set of attributes. We define an event as a data structure that consists of all the measured attributes. We use the event structure as stream element, the basic communication unit between the detector and the event classification program. The event streams are indicated by the small boxes in Figure 5 and the abstraction of its structure is superimposed. To handle the high throughput of the events at runtime, the event classification program is distributed among a group of MPI processes. Collectively, these compute processes classify the events streamed out from the detector on-the-fly. When an event arrives on a classification program, it will be discarded if it is classified as background event or saved for future analysis if it is classified as signal event. The filtered signal events will be saved to the file system during the runtime at a frequency chosen by the user.

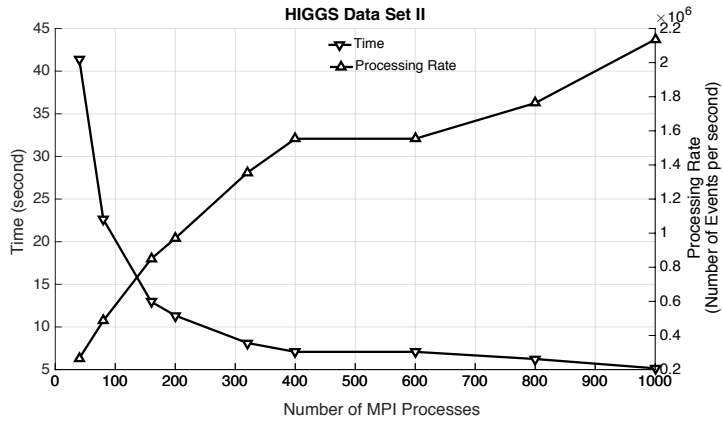

**Figure 6.** Strong Scaling of Processing Rate of LHC data on supercomputers.

We evaluate the performance of streaming processing through a realistic dataset downloaded from the data portal of Center for Machine Learning and Intelligent Systems [26]. This dataset was generated from the official ATLAS full-detector simulation. In this test, we chose a data set consisting of 11 million events. This data set has a size of approximately 7.5 GB. Each entry in the data set is an event with measurement for 28 attributes. Each event could be a signal event in which Higgs bosons were generated or just a background event in which other particles are generated. In total, there are 5,829,123 signal events to be detected in this dataset. These signal events will be saved for future research. We setup two groups of processes, one mimicking experiment sensors (called *sensor*) and the other group classifying event data (called *classifier*). The sensor processes use MPI collective I/O to collectively read data. In real case, the experiment is ongoing and the collision events are occurring so that the sensors continue detecting and sending out new events. For this reason, the sensor processes will only read in a small number of events from the data set, stream them out to the classifier processes and then repeat the loop. We chose to simulate this more expensive loop of reading and streaming instead of a one-pass buffering of whole data set in memory because it better reflects the difficulty in production run. The classifier processes keep on receiving event streams from sensors, discard background event on-the-fly and save signal events. As the classifier processes can dedicate more memory for buffering signal events, the interaction with the file systems can be largely reduced. In this test, the classifier processes do not use collective I/O to save data as the size of signal events is relatively small. We demonstrate the effectiveness of streaming processing for high event processing rate. With merely 200 processes, we can achieve over a million events per second, which is the number of event could be detected per second. The scaling of the processing rate is presented in Figure 6 against the right side Y axis. There are two phases in scaling the process rate. First, up to 400 processes, the processing rate scales quickly from 0.2 to 1.5 million events per second. Second, from 400 to 1,000 processes, the processing rate continues increasing but at slower rate, increasing from 1.5 to 2.2 million processes. Besides the impact from contention on file system, the limited number of events could result in too few events to keep all classifier processes busy when scaling to a large number of processes. However, in production run, the incoming data flow will be much larger than the test data test and the processing can be even more efficient.

## 5. Conclusions

In this paper, we showed that MPI streams can be effectively used in HPC applications. The MPIStream library provides a lightweight approach to link MPI processes with different tasks. The library is based on four basic concepts of streaming processing (communication channel, data producer/consumer, data streams and stream operations) and six functions to send and receive stream elements. We provided an example code of two separate applications using MPIStream to exchange MPI streams. We then presented three use cases using MPI streams together with their performance,

First, we implemented MapReduce with MPI streams to move key-value pairs between map and reduce processes. We presented the performance results of using the MPI stream in a MapReduce application counting the number of occurrences of words in 102,400 poems. We processed 55 million words per second when using 1,024 map processes and 33 reduce processes on the KTH Beskow supercomputer.

Second, we coupled a plasma simulation code with an I/O and visualization code to support irregular and high-frequency I/O operations and consequent visualization. The decoupling of I/O and visualization from the main scientific application allowed for a more asynchronous execution. In addition, because collective I/O are performed on a reduced number of processes instead of all the processes, collective I/O is faster. When performing a weak scaling test we observed a 3.6X improvement (with respect to iPIC3D application performing parallel I/O on all processes) on 8,192 processes using MPI streams.

In the last use case, we mimicked the LHC experiment with *sensor* processes to read particle collision events and stream them to classifier processes that filter the collision events depending on their importance. We assumed LHC sensors producing Unix/Linux device files that are opened by MPI data producers and streamed to consumers to identify the events of interest. We assumed that streaming is continuous and with constant injection rate while in reality data originating from experiment probes is highly irregular. In the case of regular streaming, we reached a processing rate of 2.2 million events analyzed per second on 1,000 cores of the KTH Beskow supercomputer.

We showed the use of MPI streams in a real-world production code like iPIC3D demonstrating the easily deployment of MPIStream in the existing codes. The main ad-vantage of using MPI streams is that it can be added incrementally to existing MPI applications without the need of using streaming frameworks, like Spark. To implement streaming between different groups only requires a small numbers of lines of codes, i.e. approximately 100 lines of code for iPIC3D and the I/O application. In the future, we plan to extend the MapReduce and LHC example to more realistic test cases and possibly to couple to data originating from experiment devices.

Overall, we showed that MPI streams can be very valuable when streaming applications need to run on supercomputer providing relatively good performance.

## Acknowledgment

This work was funded by the European Commission through SAGE project (agreement no. 671500. http://www.sagestorage.eu/). This work was supported by the DOE Office of Science, Advanced Scientific Computing Research, under ARGO (award number 66150) and CENATE projects (award number 64386). This work used resources provided by the Swedish National Infrastructure for Computing (SNIC) at PDC.